\documentclass[preprint,authoryear,12pt]{elsarticle}
\usepackage{amssymb}

\begin{document}

\begin{frontmatter}

\title{Computational approach to multifractal music}

\author[ifj]{ P.~O\'swi\c ecimka} 
\ead{pawel.oswiecimka@ifj.edu.pl}
\author[ifj]{J.~Kwapie\'n} 
\author[ifj,agh]{I. Celi\'nska} 
\author[ifj,ur]{S.~Dro\.zd\.z} 
\author[ur]{R.~Rak}

\address[ifj]{Institute of Nuclear Physics, Polish Academy of Sciences, PL--31-342 Krak\'ow, Poland}
\address[ur]{Faculty of Mathematics and Natural Sciences, University of Rzesz\'ow, PL--35-959 Rzesz\'ow,
Poland}
\address[agh]{Faculty of Physics and Applied Computer Science, AGH University of Science and Technology, PL-30-059
Krak\'ow, Poland}

\begin{abstract}
In this work we perform a fractal analysis of 160 pieces of music belonging to six different genres. We show that the
majority of the pieces reveal characteristics that allow us to classify them as physical processes called the $1/f$
(pink) noise. However, this is not true for classical music represented here by Frederic Chopin's works and for some
jazz pieces that are much more correlated than the pink noise. We also perform a multifractal (MFDFA) analysis of these
music pieces. We show that all the pieces reveal multifractal properties. The richest multifractal structures are
observed for pop and rock music.  Also the viariably of multifractal features is best visible for popular music genres.
This can suggest that, from the multifractal perspective, classical and jazz music is much more uniform than pieces of
the most popular genres of music.
\end{abstract}


\end{frontmatter}

\section{Introduction}
Since B. Mandelbrot's ``Fractal Geometry of Nature" was published~\citep{mandelbrot82}, fractals have an enormous impact
on our perception of the surrounding world. In fact, fractal (i.e. self-similar) structures are ubiquitous in nature,
and the fractal theory itself contitutes a platform on which various fields of science, such as
biology~\citep{ivanov99,makowiec09,rosas02}, chemistry~\citep{stanley88,udovichenko02},
physics~\citep{muzy08,oswiecimka06,subramaniam08},
and economics~\citep{drozdz10,kwapien05,matia03,oswiecimka05,zhou09}, come across. This (statistical) self-similarity
concerns irregularly-shaped empirical structures (Latin word fractus means 'rough') which often elude classical methods
of data analysis. This interdisciplinary character of the applied fractal geometry is not confined only to science, but
also in art, which may be treated as some reflection of reality, some interesting fractal features might be discerned.
An example of this are the fractal properties of Jackson Pollock's paintings~\citep{taylor99} and the Zipf's law
describing literary works~\citep{kwapien10,zanette06,zipf49}. In a course of time the fractal theory encompassed also
the
multifractal theory dealing with the structures which are convolutions of different fractals. It turned out that such
structures and corresponding processes are not rare in nature and the proposed multifractal formalism allowed
researchers to introduce distinction between mono- and multifractals~\citep{halsey86}. 
Development of those intriguing theories would not have been possible, though, if there had not been substantial
progress in computer science. On the one hand, fractals - due to their structure - can easily be modelled by using
iterative methods, for which the computers are ideal tools. On the other hand, however, the multifractal analyses
require significant computing power. The result of such an analysis is identification of diverse patterns in different
subsets of data which would be impossible without modern computers.
\begin{figure}
\begin{center}
\includegraphics[scale=.4]{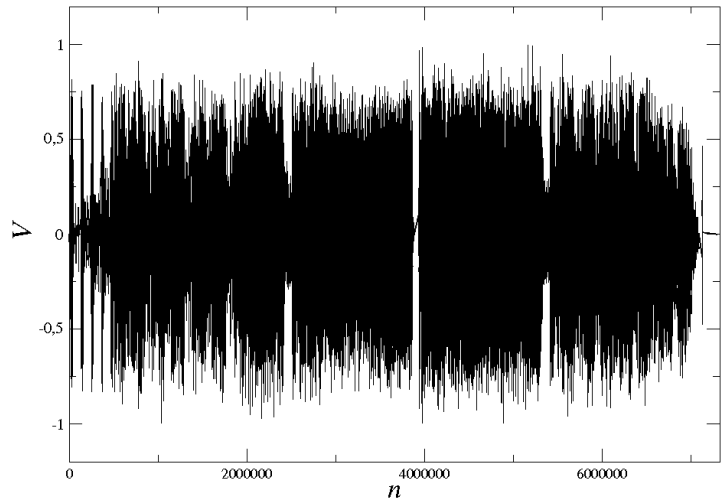}
\end{center}
\caption{Exemplary time series representing the sound wave of the song ``Good Times Bad Times" by Led Zeppelin.}
\label{serie}
\end{figure}
Although relations of mathematics and physics with music date back to ancient times (Pythagoras of Samos considered
music a science of numbers), a new impulse for them arrived together with  developments in the fractal methods of time
series analysis ~\citep{bigerelle00,ro09}. The first fractal analysis of music was carried out in 1970s by Voss and
Clarke~\citep{voss75}, who
showed that the frequency characteristics of investigated signals behave similar to $1/f$ noise. Interestingly,
this type of noise (called pink noise or scaling noise) occurs very often in nature~\citep{bak87}. The $1/f$ spectral
density is an attribute, among others, of  meteorological data series, electronic noise occurs in almost all electronic
devices, statistics of DNA sequences and heart beat rthythm~\citep{bak96}. Thus, from this point of view, music imitates
natural processes. A note worth making here is that, according to the authors of the above-cited article, the most
pleasent to ear kind of music is just the pink noise. In 1990s Hsu and Hsu showed that for some classical pieces of Bach
and Mozart and for some children songs, a power law
relation occurs between the number of subsequent notes $F$, distant from 
each other by $i$ semitones as a function of $i$~\citep{hsu90}:
\begin{equation}
 F=c/i^D
\end{equation}
where $c$ denotes a propotionality constant and $D$ is the fractal dimension $(1 < D < 2.25)$. 
\begin{figure}
\begin{center}
\includegraphics[scale=.4]{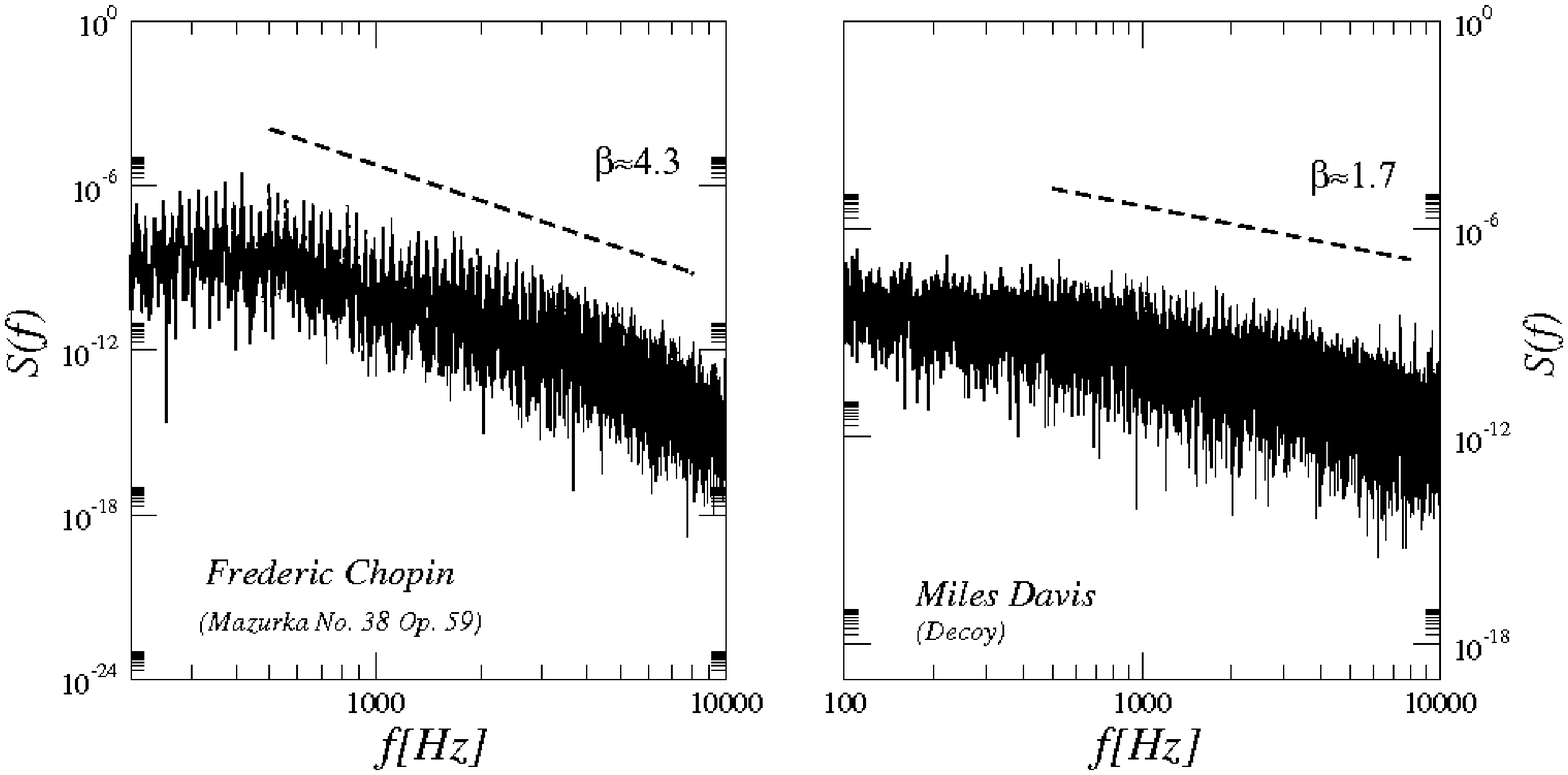}
\end{center}
\vspace{1.1cm}
\begin{center}
\includegraphics[scale=.4]{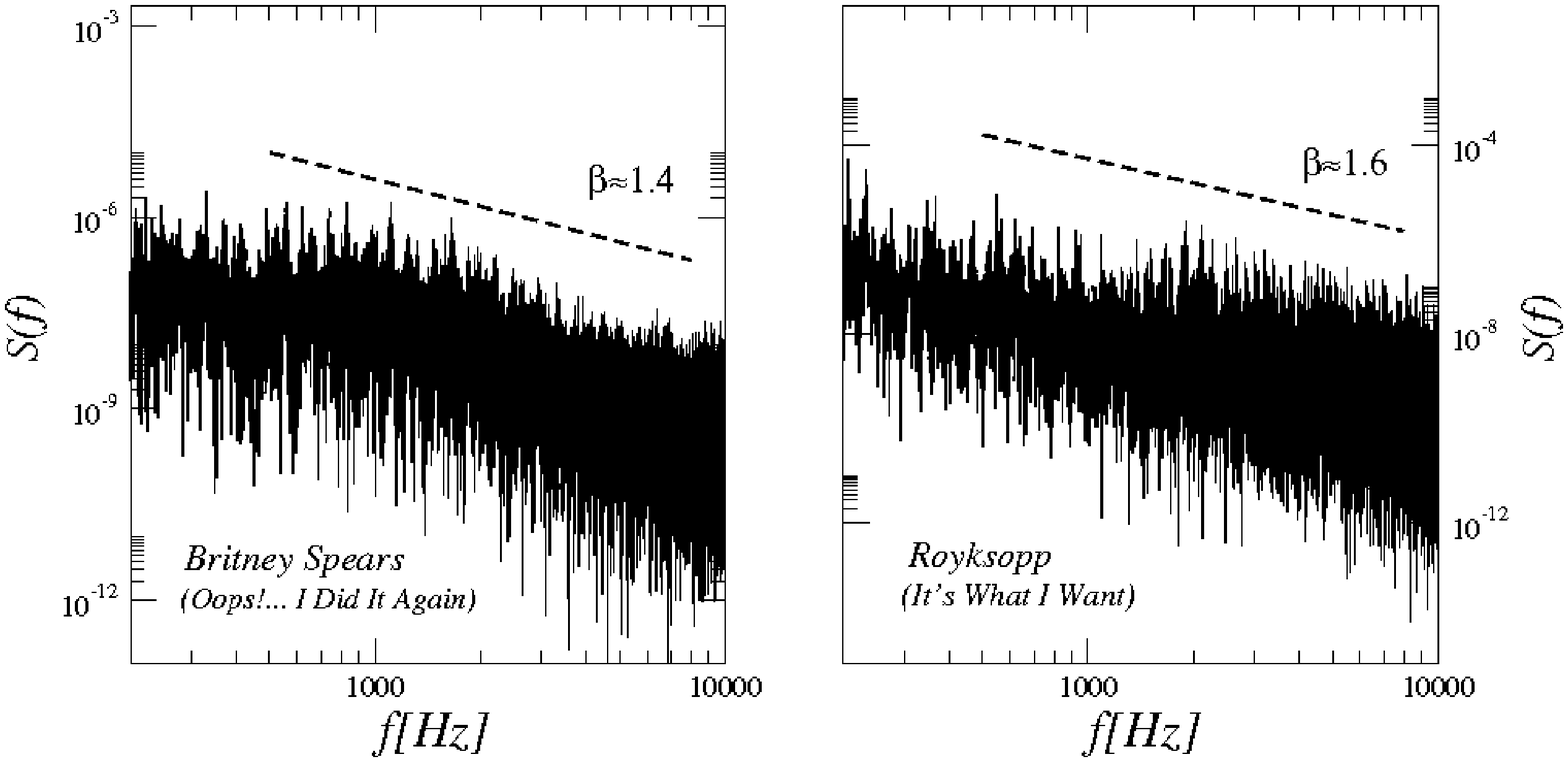}
\end{center}
\vspace{1.1cm}
\begin{center}
\includegraphics[scale=.4]{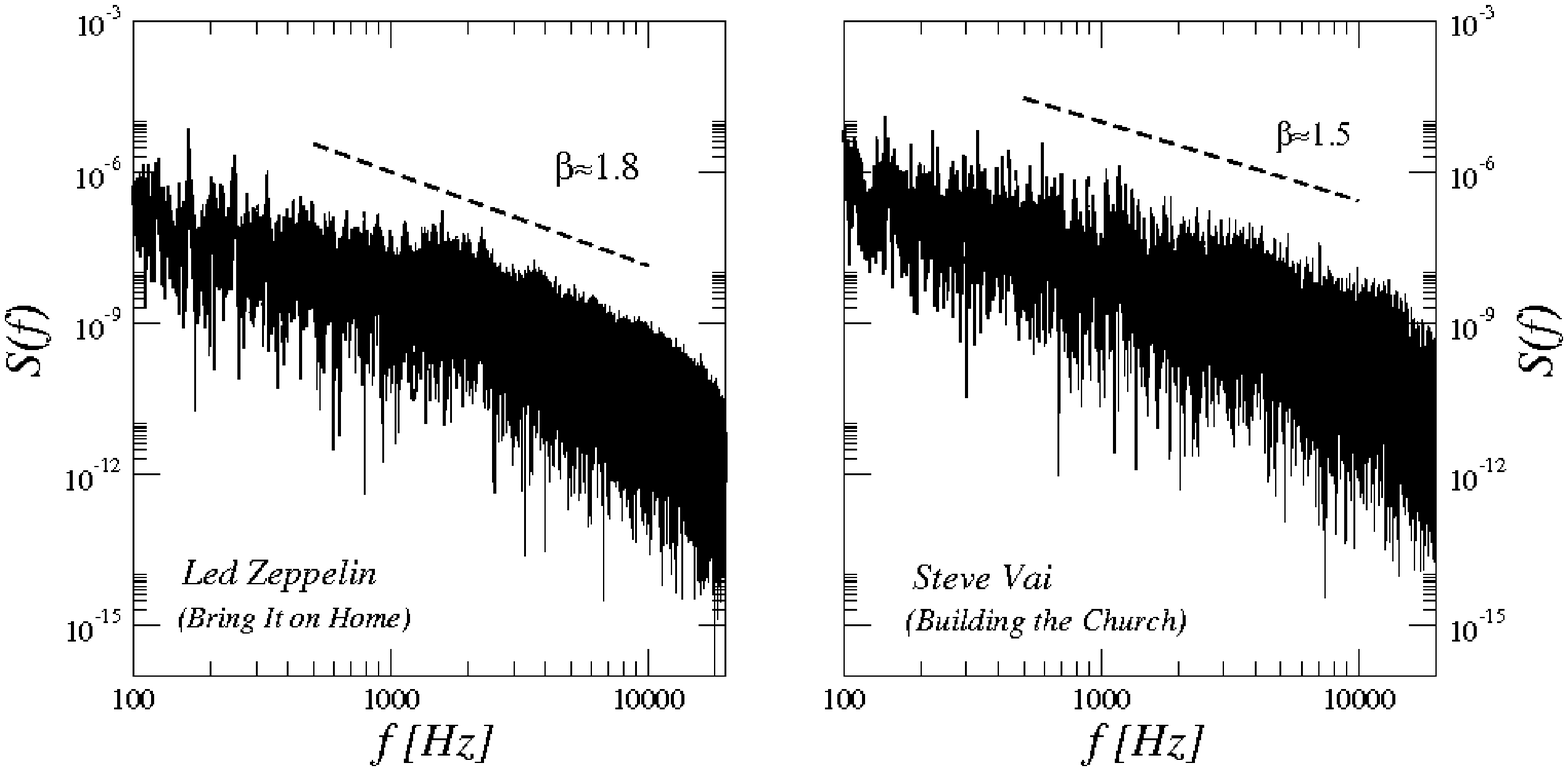}
\end{center}
\caption{ Exemplary power spectra (in log-log scale) for six pieces of music representing six genres (from top to
bottom: classical music, jazz, pop music, electronic music, rock, and hard rock). Power-law trends in each panel are
indicated by dashed lines, whose slopes correspond to the corresponding values of $\beta$.}
\label{widmo}
\end{figure}
In contrast, no similar relation has been observed for some works of Karlheinz Stockhausen, one of modern composers
belonging to the strict musical avant-garde. It should be mentioned that the relation discovered by~\cite{hsu90}. can be
considered as an expression of the Zipf's law in music. In recent years, a more advanced multifractal analysis was
carried out ~\citep{jafari07,su06}. For example, by substituting both the rhythm and melody by a geometrical sequence of
points,  ~\cite{su06} showed that these quantities can be considered the multifractal objects. They also suggested that
various genres of music may possess their genre-specific fractal properties. Thus, there might exist a multifractal
criterion for classifying a musical piece to a particular genre.
Music can be considered a set of tones or sounds ordered in a way which is pleasant to ear. And although the reception
of a musical piece is subjective, music affects a listener irrespective of his sensitivity or musical
education~\citep{storr97}. Therefore, a hypothesis which arises in this context is that music as an object may refer not
only to the structure of a musical piece but also to the way it is perceived.
\section{Power spectral analysis}
In our work we analyzed 160 pieces of music from six popular genres: classical music, pop music, rock, hard rock, jazz,
and electronic music.  The first one, classical music, was represented by 38 works by Frederick Chopin, divided into
three periods of his career. Pop music consisted of 51 songs performed by Britney Spears, rock and hard rock music - 20
songs performed by Led Zeppelin and 20 songs by Steve Vai, respectively, jazz - 25 compositions performed by Miles
Davies or Glenn Miller. Finally, an electronic music consisted of  6 pieces of music by Royskopp. All the analyzed
pieces were written in the WAV format. In this format the varying amplitude of a sound wave $V(t)$ is encoded by a
16-bit stream sampled with 44,1 kHz frequency. After encoding, the amplitude $V(t)$ was expressed by a time series of
length depending on the temporal length of a given piece of music (several million points, on average). An exemplary
time series encoding a randomly selected song is displayed in Figure \ref{serie}. We started our analysis with
calculating the power spectrum $S(f)$ for each piece of music. This quantity carries information on power density of
sound wave components of frequency $(f;df)$. According to the Wiener-Khinchin theorem, $S(f)$ is equal to the Fourier
transform of autocorrelation function or, equivalently, the squared modulus of a signal's Fourier transform:
\begin{equation}
 S(f)=|X(f)|^2
\end{equation}
where
\begin{equation}
 X(f)=\int\limits_{-\infty}^{\infty} x(t) e^{-2 \pi i ft}dt
\end{equation}
is the Fourier transform of  a signal $x(t)$. If the power spectrum decreases with $f$ as $1/f^\beta $, ($\beta\geq
0$), it means that the signal under study is characterized by log-range autocorrelation within the scales described by
the corresponding frequencies $f$. The faster is the decrease of $S(f)$ (i.e. the higher value of $\beta$), the stronger
is the autocorrelation of the signal. It is worth recalling here that the Brownian motion corresponds to $\beta=2$,
while the white noise (uncorrelated signal) to $\beta=0$. Since the exponent $\beta$ can easily be transformed into the
Hurst exponent (a well-known notion in fractal analysis) or into the fractal dimension, the power spectrum can be
classified among the monofractal techniques of data analysis. The power spectra were calculated for each piece of music.
In most cases, the graph $S(f)$ was a power-law decreasing function for frequencies 0.1-10 kHz with the slope
characteristic for a given piece. The notable exceptions were works of Chopin for which the graphs were scaling between
1 and 10 kHz. Six representative spectra for different genres are shown in Figure \ref{widmo}. To each empirical
spectrum $S(f)$ a power function was fitted (the straight lines in Figure \ref{widmo} ) within the observed
corresponding power-law regime. 
\begin{figure}
\begin{center}
\includegraphics[scale=.4]{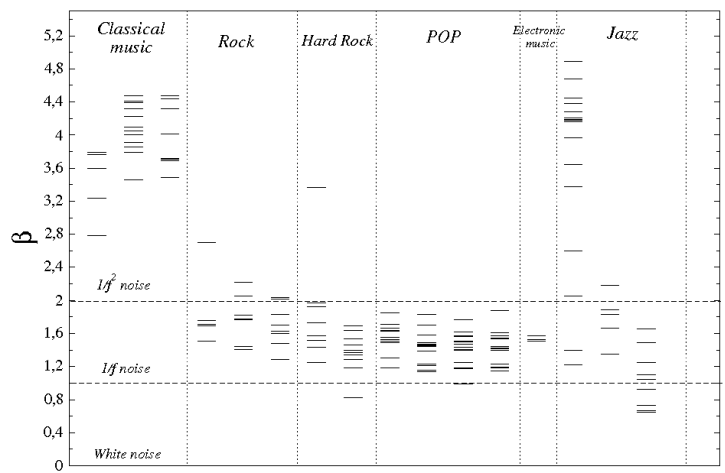}
\end{center}
\caption{Exponent $\beta$ calculated for each piece of music analyzed in the present work (short horizontal lines).
Columns correspond to individual artists, periods of their career (Chopin), or albums. Dotted vertical lines separate
different music genres.}
\label{beta}
\end{figure}
A slope of the fitted function corresponds to the exponent $\beta$. All the calculated
values of $\beta$, are exhibited in   Figure \ref{beta}. As it can be seen, the highest values of $\beta$ correspond to
works of F. Chopin (classical music,$\beta_{MAX}=~4.4$) and some works of Glenn Miller (jazz,  $\beta _{MAX}=~4.8$).
Exponents in these cases are much higher than 2 which means that the underlying processes are more correlated than the
Brownian motion. Also several songs by Led Zeppelin (rock) have $\beta>2$ but not so prominent as the pieces from those
genres mentioned before. Interestingly, $\beta$ for Led Zeppelin declines with time. For their
chronologically first album, the highest exponent is 2.8, while for the subsequent albums it drops to 2.3 and 2.1,
respectively. For the other analyzed music genres, i.e. electronic, pop, rock and hard rock music, $1 < \beta < 2$. It
is also worth mentioning that for several jazz pieces, the exponent $\beta$ drops below 1, which means  that they
approach white noise. An author of these songs is Miles Davies, one of the most significant jazz artists, who often was
a precursor of new styles and sounds. To summarize this part of our analysis, we can say that from the power spectrum
perspective, the majority of the analyzed pieces of music can in fact be considered the $1/f$ processes. This is even
more evident for more popular music genres like pop and rock than for rather exclusive genres like jazz and classical
music. 
\section{Multifractal analysis of musical compositions}
In order to have a deeper insight into dynamics of the investigated signals we performed also multifractal
analysis of data. We used one of the most popular and reliable methods - the Multifractal Detrended Fluctuation Analysis
(MFDFA)~\citep{kantelhardt02}. This method allows us to calculate fractal dimensions and Hoelder exponents for
individual components of a signal decomposed with respect to the size of fluctuations. Consecutive steps of this
procedure are presented below. At the beginning we calculate the so-called profile, which is the cumulative sum of the
analyzed signal:
\begin{equation}
Y(i)=\sum \limits_{j=1}^{i}[x_j-\langle x\rangle] \quad \hbox{for} \quad i=1,2,...N ,       
\end{equation}
where $\langle x\rangle$ donotes the signal's mean, and $N$ denotes its length. The subtraction of the mean value is not
necessary, because a trend is eliminated in the next steps. Then we divide the profile $Y(i)$ into $N_s$ disjoint
segments of lentgh $s$ ($N_s=N/s$). In order to take into account all the points (at the end of the signal's profile
some data can be neglected), the dividing procedure has to be repeated starting from the end of $Y(i)$. In consequence,
we obtain $2N_s$ segments. In each segment $\nu$, the estimated trend is subtracted from the data. The trend is
represented by a polynomial $P_{\nu}^{l}$ of order $l$. The polynomial's order used in calculation determines the
variant of the method. Thus, for $l=1$ we have MFDFA1, for $l=2$ - MFDFA2, and so on. After detrending the data, its
variance has to be calculated in each segment:
\begin{figure}
\begin{center}
\includegraphics[scale=.4]{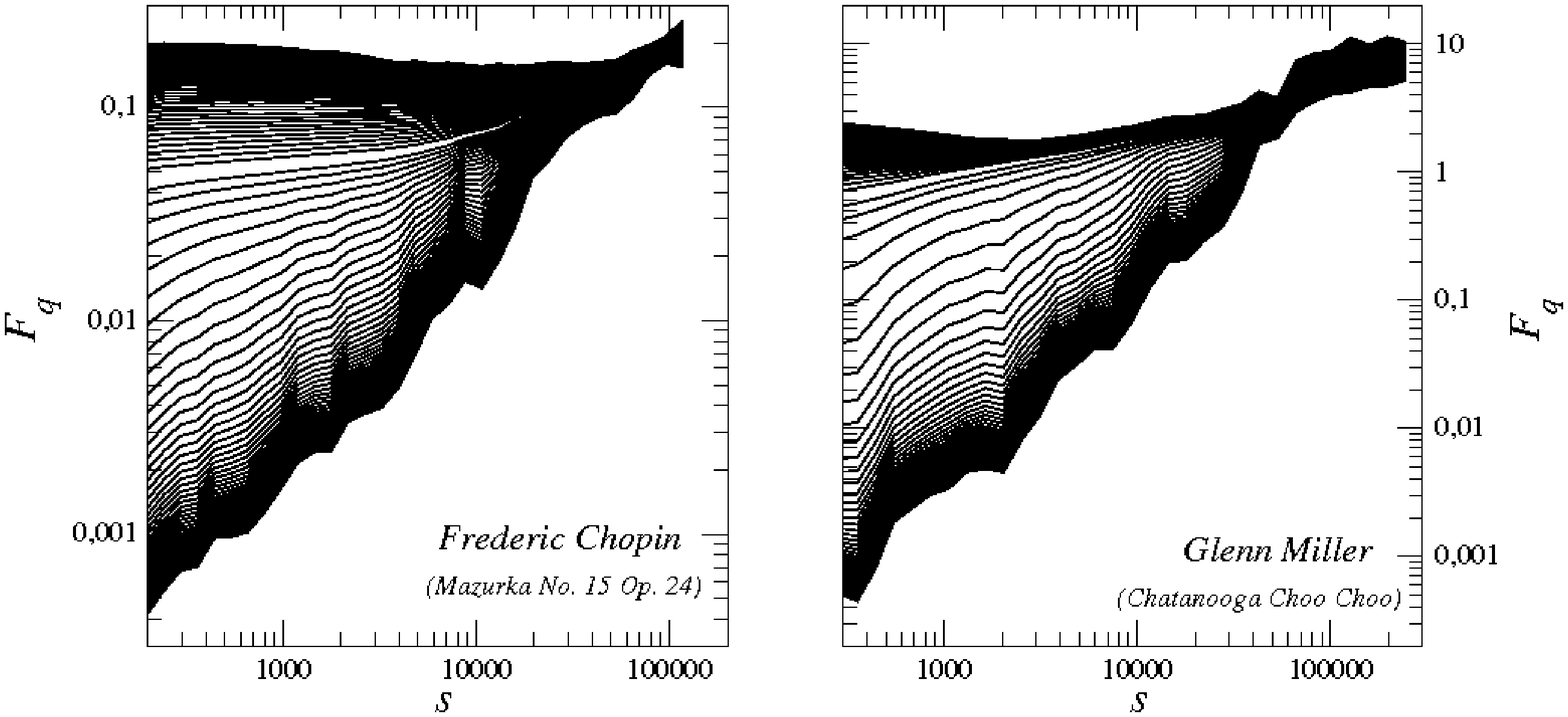}
\end{center}
\vspace{1.1cm}
\begin{center}
\includegraphics[scale=.4]{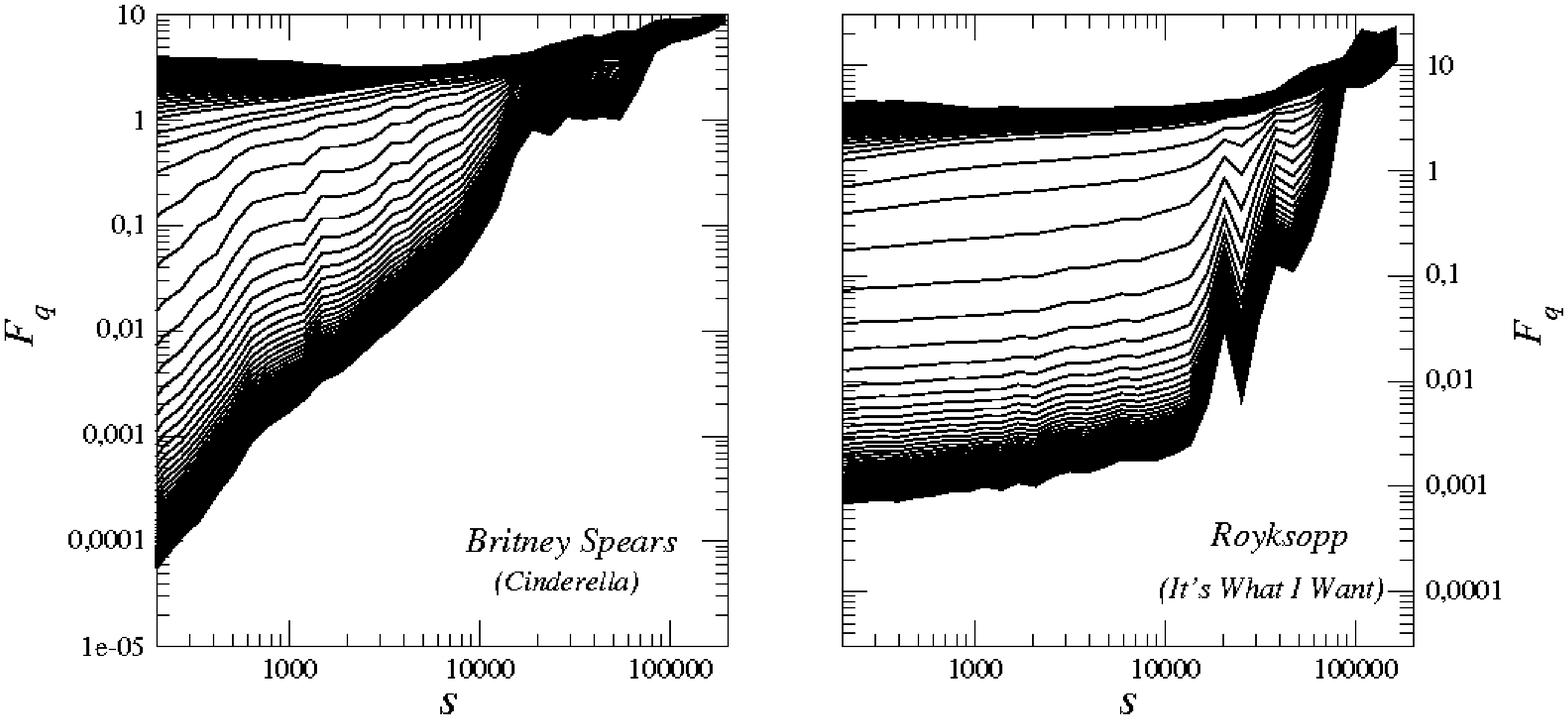}
\end{center}
\vspace{1.1cm}
\begin{center}
\includegraphics[scale=.4]{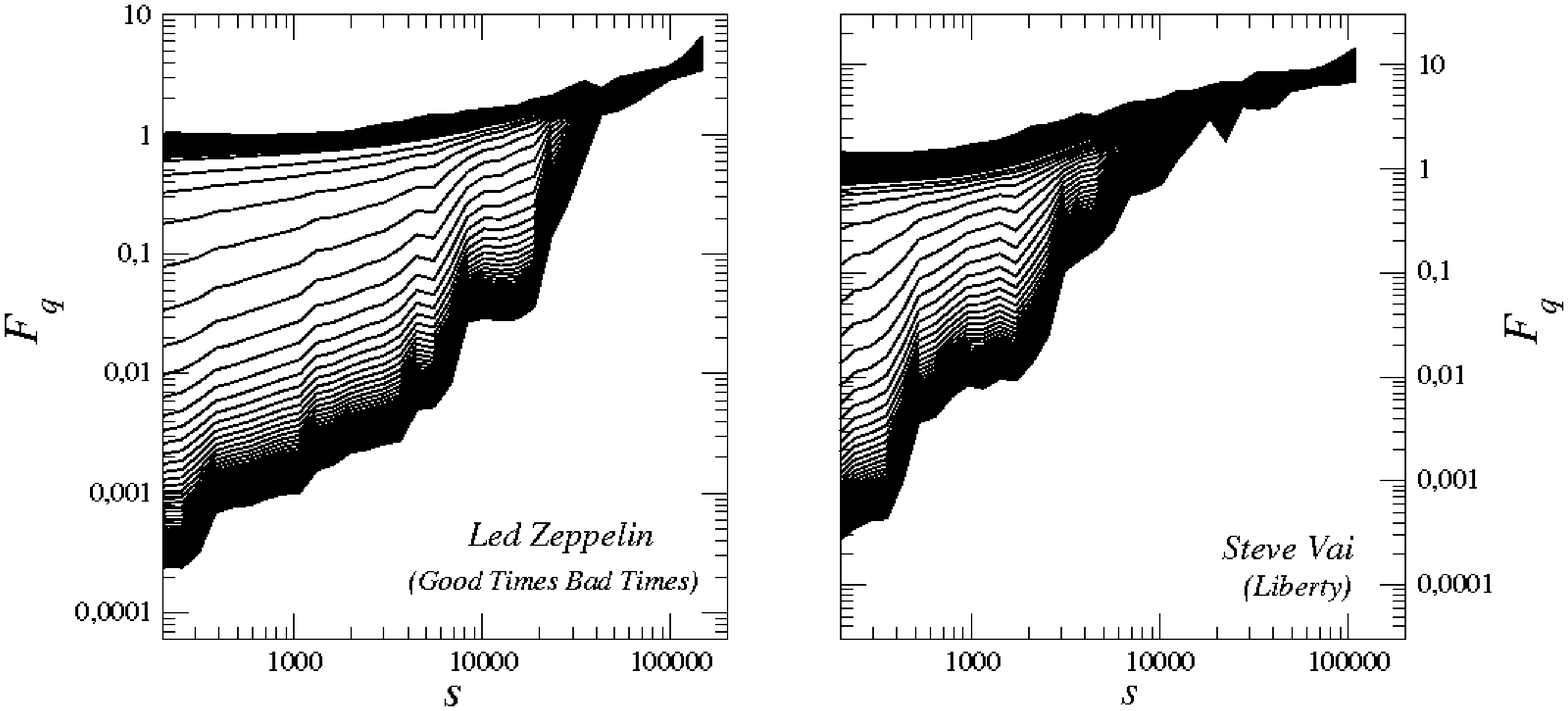}
\end{center}
\caption{ Exemplary fluctuation function $Fq$ (in $log-log$ scale) for six pieces of music representing six 
genres (from top to bottom: classical music, jazz, pop music, electronic music, rock and hard rock). Each line
represents
$Fq$ calculated for particular $q$ values in the range from -4 to 4}
\label{Fq}
\end{figure}
\begin{equation}
 F^2(s,\nu)=\frac{1}{s}\sum \limits_{i=1}^{s}\{Y[(\nu -1)s+i]-P_{\nu}^{l}(i)\}^2 \quad \hbox{for} \quad \nu=1,2,...N_s
\end{equation}
 or
\begin{equation}
 F^2(s,\nu)=\frac{1}{s}\sum \limits_{i=1}^{s}\{Y[N-(\nu -N_s)s+i]-P_{\nu}^{l}(i)\}^2 \quad \hbox{for} \quad
\nu=N_s+1,N_s+2,...2N_s
\end{equation}  
The variances are then averaged over all the segments and, finally, one gets the $q$th order fluctuation function given
by:
\begin{equation}
 F_q(s)=\left \{ \frac{1}{2N_s}\sum \limits_{\nu =1}^{2N_s} [F^2(s,\nu)]^{q/2} \right \}^{1/q} ,
\end{equation}
where the exponent $q$ belongs to $\mathbb{R} \setminus \{ 0 \}$. This procedure has to be repeated for different values
of $s$. If the analyzed signal has any fractal properties, the fluctuation function behaves as:
\begin{equation}
 F_q(s)\sim s^{h(q)},
\end{equation}
where $h(q)$ denotes the generalized Hurst exponents. 
A constant $h(q)$ for all $q$'s means that the studied signal is
monofractal and $h(q)=H$ (the ordinary Hurst exponent). For multifractal signals, $h(q)$ is a monotonically decreasing
function of $q$. It can be easily noticed that, by varying the $q$ parameter, it is possible to decompose the time
series into fluctuation components of different character: for $q > 0$ the fluctuation function mostly describes large
fluctuations, whereas for $q < 0$ the main contribution to the $F_q$ comes from small fluctuations. By knowing the
$h(q)$ spectrum for a given set of data, we can calculate its singularity (multifractal) spectrum:
\begin{equation}
 \alpha = h(q)+qh^{'}(q) \quad \hbox{and} \quad f(\alpha)=q[\alpha-h(q)]+1 ,
\end{equation}
where $h^{'}(q)$ stands for the derivative of $h(q)$ with respect to $q$, the Hoelder exponent $\alpha$ donotes
singularity strength, and $f(\alpha)$ is the fractal dimension of the set of points characterized by $\alpha$. For a
monofractal time series, the singularity spectrum reduces to a single point $(H,1)$, while for multifractal
time series, the spectrum assumes the shape of an inverted parabola. The multifractal strength is a quantity which
describes the richness of multifractality, i.e., how diverse are values of the Hoelder exponents in a data set. It can
be estimated by the width of the $f(\alpha)$ parabola:
\begin{equation}
 \Delta \alpha = \alpha _{max} - \alpha _{min}
\end{equation}
\begin{figure}
\begin{center}
\includegraphics[scale=.4]{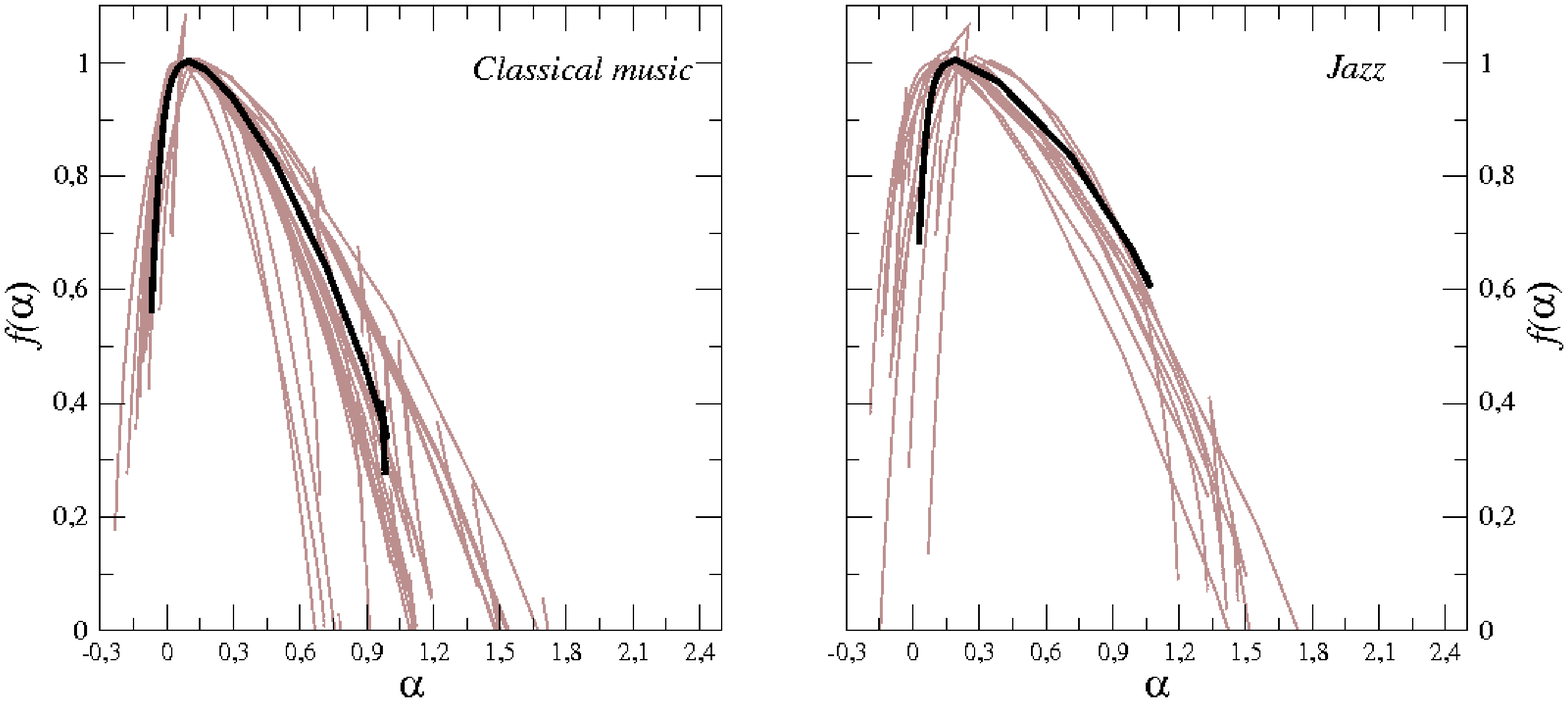}
\end{center}
\vspace{1.1cm}
\begin{center}
\includegraphics[scale=.4]{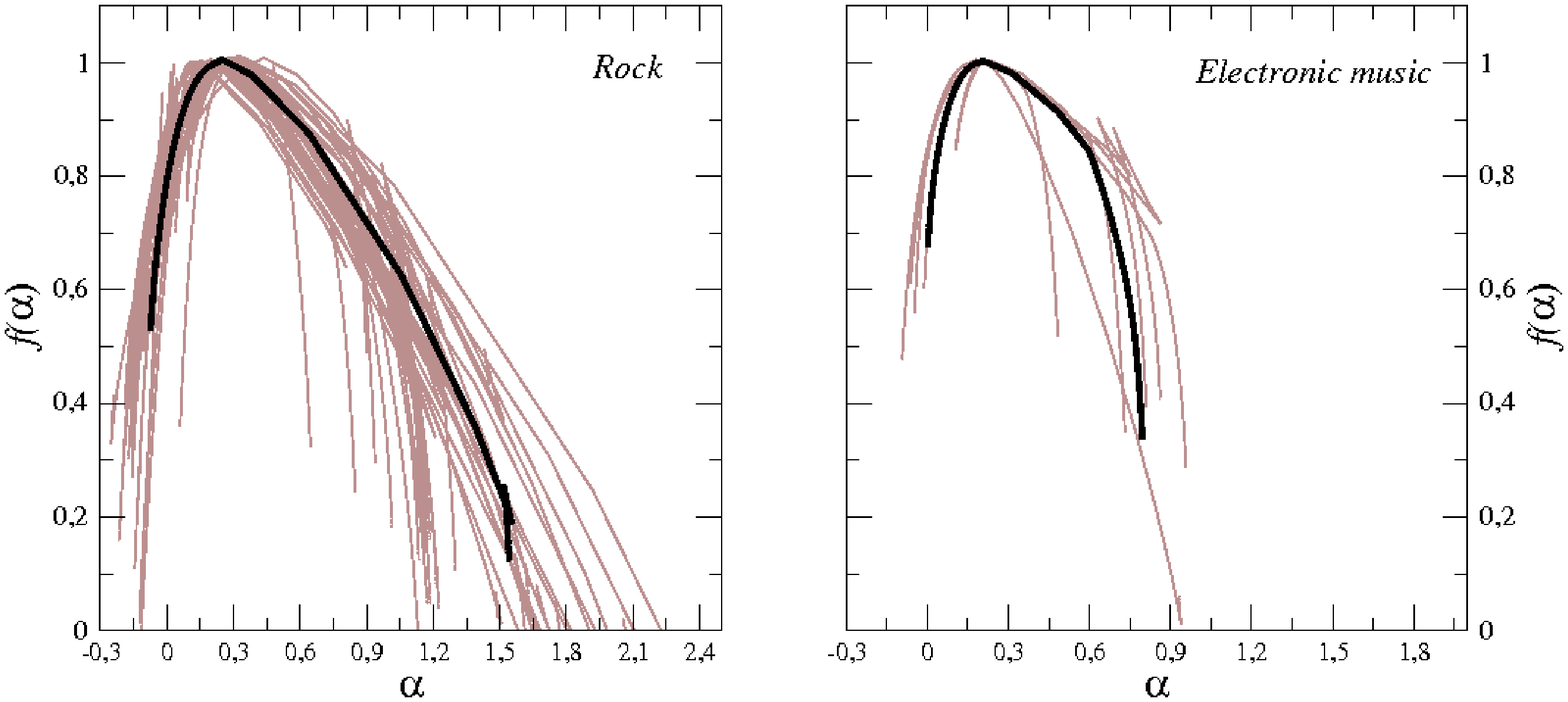}
\end{center}
\vspace{1.1cm}
\begin{center}
\includegraphics[scale=.4]{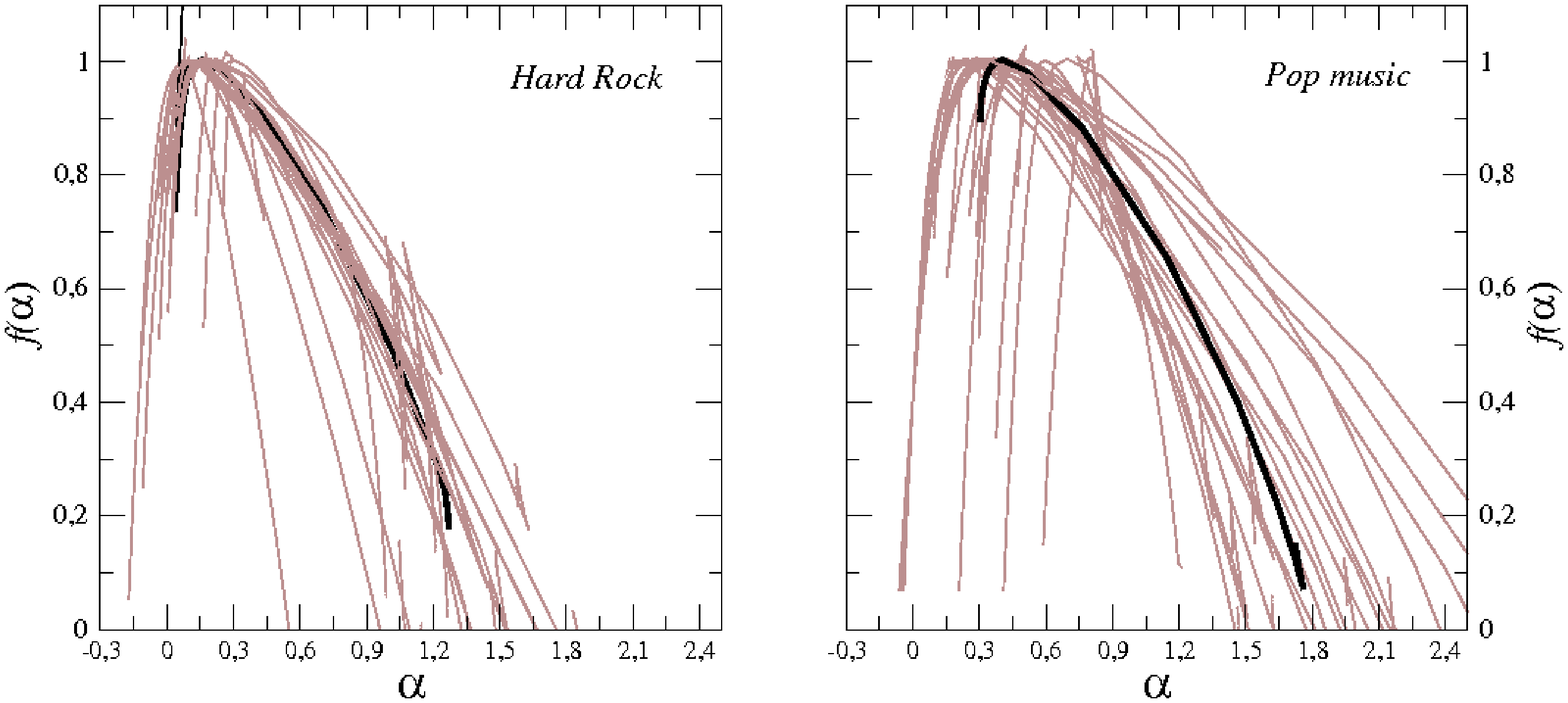}
\end{center}
\caption{The multifractal spectra $f(\alpha)$ calculated for each analyzed piece of music (brown lines) and its mean
spectrum (black lines).}
\label{falfa}
\end{figure}
where $\alpha _{min}$ and $\alpha _{max}$  stand for the extreme values of $\alpha$. The bigger is $\alpha$ the richer
is the multifractal. \newline
Using MFDFA2, which guarantees stability of results, we calculated the fluctuation function $Fq$ for all the analyzed
signals in the scale s range from 50 to 100,000 points. The value of $q$ was increased by 0.2 in the range from -4 to 4.
Exemplary fluctuation functions for our six music genres are shown in Figure \ref{Fq}. All the calculated $Fq$ functions
are characterized by a power law dependence on the scale for all $q$'s. However, the range of scaling  varies slightly
for different pieces. By looking at the shown examples, it is easy to notice that for F. Chopin, Britney Spears, Glenn
Miller, and Steve Vai, the scaling involves almost all the considered values of $s$, while for electronic music we can
distinguish two scaling ranges: the longer one for the scales $40 < s < 10,000$ and the shorter one for $10,000 < s <
100,000$. Such double scaling appears also occasionally for the other genres of music. However, for the most cases, we
observe only one type of scaling. In Figure \ref{Fq} we can also notice a clear dependence of the $h(q)$ exponent (the
slope coefficient of $Fq$ in double log scale) on $q$. And so, the largest values of $h(q)$ correspond to $q < 0$,
whereas for $q > 0$, $h(q)$ takes smaller values. Therefore already at this stage of calculations, it can be seen that
the analyzed signals can have distinct multifractal properties.  It is also worth to mention that for the large scales
(e.g. for jazz $s > 40,000$, for hard rock $s > 20,000$), scaling loses its multifractal traits, and $h(q)$ does not
depend on $q$. It is related to the limited range of nonlinear correlations~\citep{drozdz09}. The scale $s$, for which
the scaling character of $Fq$ changes, sets a limit for estimation of the multifractal spectrum.
For all the fluctuation functions, we estimated the singularity spectra $f(\alpha)$. Figure \ref{falfa} presents the
multifractal spectra (grey lines) and the corresponding mean multifractal spectra (black lines) for the music genres to
which the given pieces belong. All the mean spectra are asymmetric. The right part which describe the small amplitude
fluctuations is clearly longer. This effect is best visible for the rock, hard rock, and pop songs. Locations of the
extrema of these spectra ($\alpha\approx 0.2$) suggest considerably antipersistent behavior of the analyzed time series.
We can easily see that the width of the multifractal spectra for a particular genre fluctuates considerably.
Nevert heless, all the spectra are characterized by the widths large enough that they can be regarded as multifractal
structures. This confirms observation made above for the $Fq$  function. The narrowest mean multifractal spectrum was
observed for electronic music ($\Delta\alpha= 0.85$). Classical music and jazz display mutually comparable widths of,
respectively, 1.0 and 1.1. The widest mean $f(\alpha)$ is seen for hard rock (1.22), rock (1.5), and pop (1.8). 
\begin{figure}
\begin{center}
\includegraphics[scale=.4]{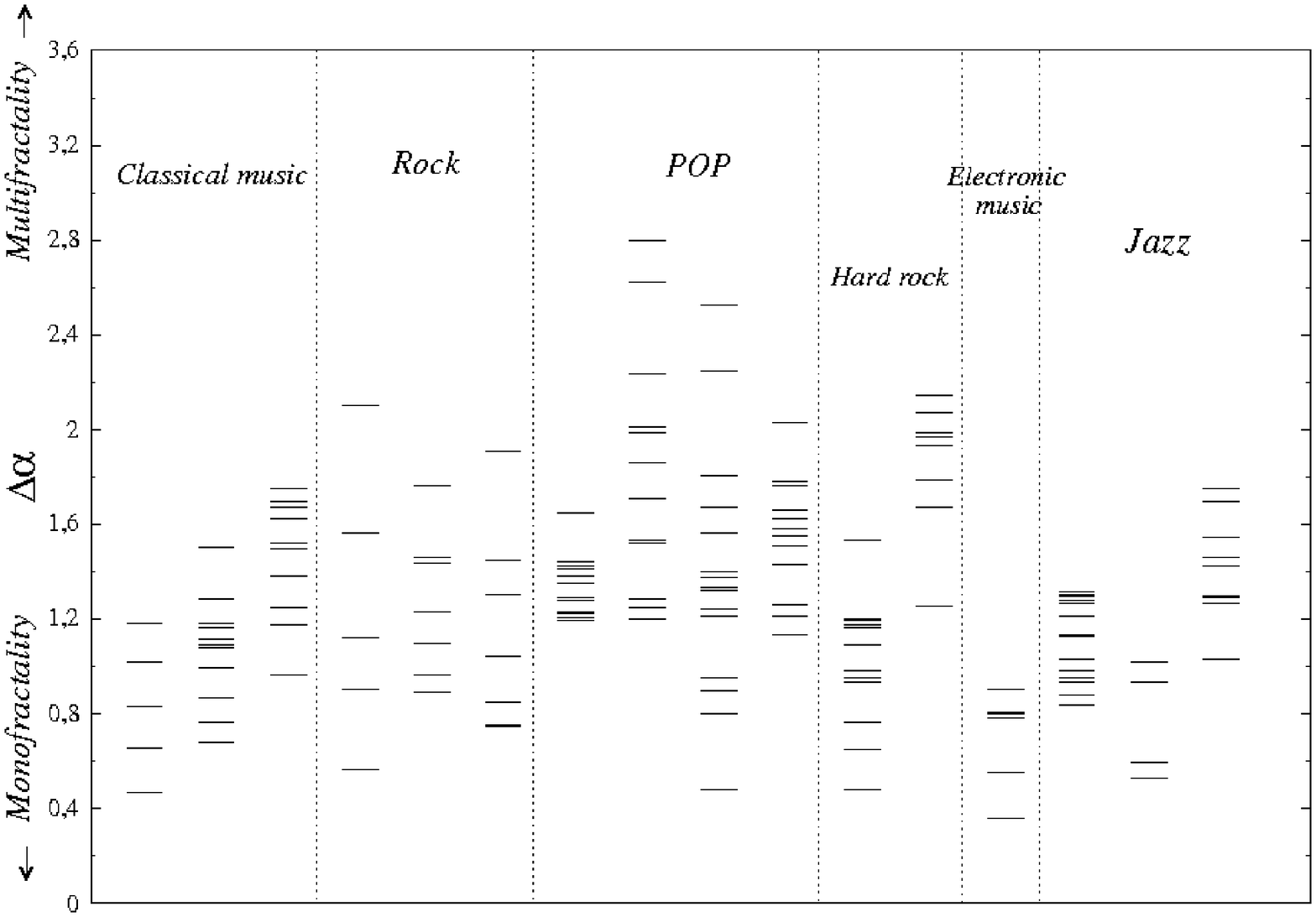}
\end{center}
\caption{Value of $\Delta \alpha$ calculated for each  analyzed piece of music (short horizontal lines). Columns
correspond to individual artist, periods of their career, or albums. Dotted vertical lines separate different genres of
music.}
\label{dalfa}
\end{figure}
Thus, from this point of view, the richest multifractal (the richest dynamics of processes) is an attribute of the most
popular music genres. The more exclusive genres are characterized by poorer multifractals. Figure \ref{dalfa} presents a
collection of all the calculated widths of $f(\alpha)$. Vertical lines separate different music genres and each piece is
represented by a single horizontal line. As it can be seen, the most variable multifractal spectra widths characterize
pop ($0.5 <\alpha< 2.8$), rock  ($0.5 < \alpha< 2.1$) and hard rock ($0.51 < \alpha < 2.15$) music. Thus, on account of
multifractal properties, the pieces belonging to these genres differ markedly among themselves. Much more consistent
from this point of view are the pieces of classical music, jazz and electronic music. We can draw therefore a conclusion
that this is the richness of multifractal forms what distinguishes popular music from the more exclusive and the less
listened to musical genres.
\section{Conclusions}
To sum up, our work presents results of a fractal analysis of selected music works belonging to six different genres:
pop, rock, hard rock, jazz, classical, and electronic music. The results confirm that the amplitude signals $V(t)$ are
characterized by the power spectrum falling off according to a power law: $S(f)\sim 1/f^\beta $. Interestingly, rate
of this fall can be characteristic for a particular genre. For classical music and some pieces of jazz, $S(f)$ declines
the fastest, while for popular music (pop, rock, hard rock , and electronic music ) the power spectrum falls more slowly
suggesting less correlated signals. The same signals were also subject to a multifractal analysis. It turned out that
such data demonstrate well-developed multifractality. Interestingly, the most variable widths of multifractal spectra
(and also the widest singularity spectra thus strongest nonlinear correlations) were observed for popular genres like
pop and rock. For the remaining genres, the multifractal properties were rather similar among the pieces. Therefore,
from this point of view, the popular music is characterized by the amplitude signals with different degree of
correlations, whereas more sophisticated musical genres (classical, jazz) are more consistent in this matter.

\end{document}